\documentclass[aps,pra,nofootinbib,superscriptaddress,twocolumn]{revtex4-1}

\usepackage[utf8,latin1]{inputenc}
\usepackage[T1]{fontenc}     
\usepackage[british]{babel}
\usepackage[dvipsnames]{xcolor}
\usepackage[colorlinks=true,citecolor=Green,linkcolor=Red,urlcolor=Cyan,hyperindex]{hyperref}
\usepackage{graphicx}
\usepackage{epsfig}
\usepackage{lipsum}     
\usepackage{color}
\usepackage{lmodern}
\usepackage[babel=true]{microtype}
\usepackage{amsmath,amssymb,amsthm}
\usepackage[capitalize]{cleveref}
\usepackage{mathtools}
\usepackage{bm}
\usepackage{soul}
\usepackage{enumerate}
\usepackage{natbib}
\usepackage[toc,page]{appendix}
\newtheorem{thm}{Theorem}
\newtheorem{lemma}{Lemma}

\newcommand{\ket}[1]{\vert#1\rangle}
\newcommand{\norm}[1]{||#1||}
\newcommand{\bra}[1]{\langle#1\vert}

\newcommand{\tr}{\text{\normalfont Tr}}

\newcommand{\II}{\mathbb{I}}

\allowdisplaybreaks

\begin{document}

\title{Quasiprobabilistic state-overlap estimator for NISQ devices}

\author{Leonardo Guerini}
\affiliation{Department of Mathematics, Federal University of Santa Maria, Santa Maria, RS 97105-900, Brazil}
\affiliation{Institute of Physics, Federal University of Rio de Janeiro, Rio de Janeiro, RJ 21941-972, Brazil}

\author{Roeland Wiersema}
\affiliation{Vector Institute for Artificial Intelligence, MaRS Centre, Toronto, ON, Canada M5G 1M1}

\author{Juan Felipe Carrasquilla}
\affiliation{Vector Institute for Artificial Intelligence, MaRS Centre, Toronto, ON, Canada M5G 1M1}

\author{Leandro Aolita}
\affiliation{Institute of Physics, Federal University of Rio de Janeiro, Rio de Janeiro, RJ 21941-972, Brazil}
\affiliation{Quantum Research Centre, Technology Innovation Institute, Abu Dhabi, United Arab Emirates}

\begin{abstract}
As quantum technologies mature, the development of tools for benchmarking their ability to prepare and manipulate complex quantum states becomes increasingly necessary.
A key concept, the state overlap between two quantum states, offers a natural tool to verify and cross-validate quantum simulators and quantum computers.
Recent progress in controlling and measuring large quantum systems has motivated the development of state overlap estimators of varying efficiency and experimental complexity.
Here, we demonstrate a practical approach for measuring the overlap between quantum states based on a factorable quasiprobabilistic representation of the states, and
compare it with methods based on randomised measurements. Assuming realistic noisy intermediate scale quantum (NISQ) devices limitations, our quasiprobabilistic method outperforms the best circuits designed for state-overlap estimation for n-qubit states, with n > 2.
For n < 7, our technique outperforms also the currently best direct estimator based on randomised local measurements, thus
establishing a niche of optimality.
\end{abstract}

\maketitle


The SWAP test, also known as the ``equality algorithm'', is the gold-standard primitive for estimating the overlap between two unknown quantum states directly (i.e., without expensive tomographic reconstructions).
Originally introduced in the context of communication-complexity problems\cite{Buhrman2001}, it is nowadays a key subroutine for a variety of quantum algorithms, ranging from quantum kernel-based machine learning (e.g. support vector machines, Gaussian processes, clustering, and principal component analysis\cite{Schuld21, Benedetti_2019, lloyd2013, Rebentrost_2014, Zhao_2019}), to training deep quantum neural networks\cite{Beer_2020} (e.g. for state or circuit learning), or validation of quantum devices.


Although conceptually simple, the SWAP test is unpractical for noisy intermediate-scale quantum (NISQ) circuits. 
In the standard approach, the test involves a SWAP gate between two $n$-qubit registers (that carry the input states) 
controlled by an ancillary qubit, i.e. an $n$-qubit Fredkin gate.
Due to limited hardware connectivity, this gate must be synthesised by a deep circuit that rapidly consumes all the available quantum-depth budget. 
E.g., for 1D circuits with nearest-neighbour connectivity, the standard way of exactly synthesising the $2$-qubit Fredkin gate requires a circuit of 212 layers.
But devoting such a large amount of quantum hardware on an operationally simple task goes against the NISQ mantra ``run easy tasks classically and save the quantum hardware for the actually hard subroutines''.

In this spirit, important simplifications have been attained. For instance, a classical machine-learning approach\cite{Cincio2018} has recently revealed an alternative exact synthesis of the $n$-qubit Fredkin gate using fewer gates.
Additionally, an even shallower circuit for the state overlap has been found that exploits transversal Bell measurements across the $n$-qubit registers and requires no ancillary qubit\cite{Garcia2013}.
Nevertheless, these approaches are still not practical enough for NISQ devices.
In particular, synthesising Bell measurements between distant qubits with circuits of limited-connectivity still requires circuit depths that rapidly become prohibitive as $n$ increases.

In a conceptually different approach, some entanglement-free methods to direct overlap estimation\cite{da_Silva_2011, Flammia_2011} were proposed.
A recent work in this direction presents an estimator based on single-qubit random measurement choices correlated between both platforms through classical communication\cite{Elben2020}.
Nevertheless, its numerical analysis focuses mostly on the overlap of a state with itself (purity), leaving aside the more general case between different states, required in most algorithmic applications of the SWAP test.
In fact, no comparison with the aforementioned entanglement-based approaches was reported there.

In this paper we demonstrate an entanglement-free technique for direct state-overlap estimation, built upon a recently studied factorable probabilistic representation of quantum states\cite{Carrasquilla_2019}.
In similar fashion to Ref. \cite{Elben2020}, we employ single-qubit measurements, 
but require no measurement coordination between both system registers.
In addition, instead of focusing on state purities, we investigate general overlaps between arbitrary state pairs.
For $ n $-qubit states, with $ n<7 $, we find our approach to present the lowest mean absolute error. 
We also compare our technique against circuit-based approaches,
that classically simulate noisy implementations of two different SWAP test variants.
In this context, we analyse the average statistical error of our method against their total average error (statistical plus systematic due to gate error accumulation). 
For typical noise parameters of current hardware implementations, our method considerably outperforms (in terms of mean absolute error) the ancilla-qubit based SWAP tests already for all $n\geq 2$ and the Bell-measurement variant for all $n\geq3$, while the sample complexity remains well below that of full state tomography or even quantum compressed sensing\cite{Gross_2010, Riofr_o_2017}.

The method we present here is also related to estimation from classical shadows\cite{Huang_2020}, an efficient approach to approximate linear functions of quantum systems without fully characterising them.
In particular, the shadow fidelity estimator probes a given state and compare it with a classical description of a target state.
Although we sample both (unknown) states, we can see both methods as different variants of the same overall strategy.
In fact, particular choices in our construction leads to a particular instance of their estimator, and vice-versa.

Our numerical results rely on an efficient tensor-network construction (based on locally-purified density operators\cite{Werner_2016}) tailored for non-unitary dynamics.
This allows for realistic simulations of multiqubit gate noise in quantum circuits.

\section{Overlap estimation via a quasi-probabilistic representation}

A $ n $-qubit factorable, informationally complete positive-operator-valued measure (IC-POVM) is a set of operators $ \{M_{a}^{(n)}=M_{a_1}\otimes\ldots\otimes M_{a_n}\}_a $.
This means the POVM elements are positive semi-definite operators $ M_a^{(n)}\geq0 $ from $ \mathcal{L}(\mathbb{C}^2\otimes\ldots\otimes\mathbb{C}^2) $ that satisfy $ \sum_a M_a^{(n)} = \II $ and span the space of Hermitian operators.
This POVM defines an associated matrix $T = [\tr(M_a^{(n)} M_b^{(n)})]_{ab}$.
Information-completeness allows us to write any $ n $-qubit state $ \rho $ as\cite{Carrasquilla_2019}
\begin{equation}\label{quasi_repr}
\rho = \sum_{a, a'} P_{\rho}(a)\tau_{aa'}M_{a'}^{(n)},
\end{equation}
where $ P_\rho(a) = \tr(\rho M_a^{(n)}) $ and $ \tau$ can be defined as the pseudoinverse of $ T $ (however, as we show below, other choices of $ \tau $ are possible).
Since the POVM elements are not orthogonal, the associated probabilities do not refer to exclusive events, and therefore $ P_\rho $ is called a quasiprobability.

Let us now proceed to characterise all choices of $ \tau $.
Noticing that decomposition (\ref{quasi_repr}) holds for any operator, we can apply it to $ \rho = M_b^{(n)} $ and obtain
\begin{align}\nonumber
[T]_{bb'} =& \tr(M_{b}^{(n)}M_{b'}^{(n)}) \\ =& \sum_{a, a'}\tr(M_b^{(n)}M_a^{(n)})\tau_{aa'}\tr(M_{a'}^{(n)}M_{b'}^{(n)}),
\end{align}
for any $b, b'$.
This yields the relation
\begin{equation}\label{tau}
T = T\tau T.
\end{equation}
Any matrix satisfying (\ref{tau}) is called a generalised inverse for $ T $; in particular, the pseudoinverse fulfils this condition.
Analogously, again from the completeness of the POVM, it is straightforward to check that (\ref{tau}) conversely implies (\ref{quasi_repr}) (see Appendix \ref{app:lemma}).
This proves the following observation.

\begin{lemma}\label{lemma}
	Eq. (\ref{quasi_repr}) holds for any state $ \rho $ if and only if $ T=T\tau T $.
\end{lemma}

\begin{figure}
	\includegraphics[width=\linewidth]{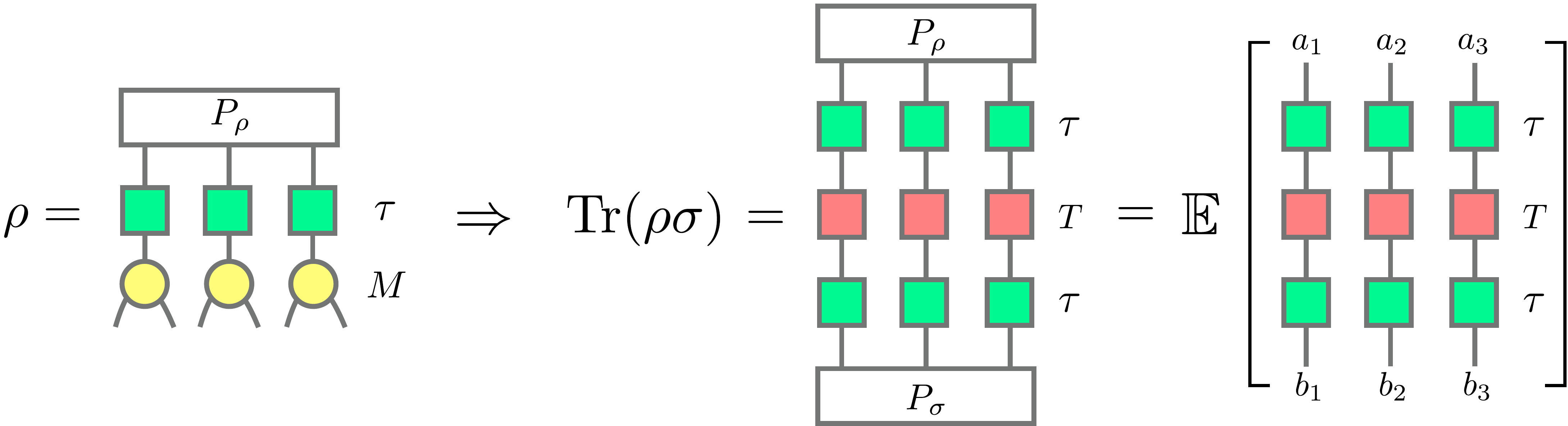}
	\caption{Tensor network representing our state-overlap estimator, based on a quasi-probabilistic representation for quantum states via an IC-POVM $ M $, its associated matrix $ T $ and an auxiliary tensor $ \tau $.}
	\label{fig:estimator}
\end{figure}

Notice that we can use different tensors $ \tau_1,\tau_2 $ for representing different quantum states.
Using this fact to represent states $ \rho $ and $ \sigma $, we find their overlap is given by 
\begin{align}\label{estimator}\nonumber
\tr(\rho \sigma) =& \sum_{a,b}P_\rho(a)[\tau_1T\tau_2^t]_{ab}P_\sigma(b) \\
=& \underset{a\sim P_\rho, b\sim P_\sigma}{\mathbb{E}}[\tau_1T\tau_2^t]_{ab}.
\end{align}
This provides a practical way of estimating the overlap Tr$ (\rho\sigma) $.
First, we repeatedly measure $ M^{(n)} $ on $ \rho $ and $ \sigma $.
This can be done separately and at different times.
Then, according to the obtained outcomes $ a,b $, we approximate the above expectation value by an average of the matrix elements $ \hat\tau=[\tau_1T\tau_2^t]_{ab} $ (see Fig. \ref{fig:estimator}).

This approximation generates a statistical error, which we now consider.
Recall that representation (\ref{quasi_repr}) allows to reconstruct entangled states $ \rho $ as combinations of the separable operators $ M_a $, implying that $ \hat\tau$ forcibly has negative entries.
This negativity can be quantified in terms of the range $ \nu = \max_{a,b}[\hat\tau]_{ab} - \min_{a,b}[\hat\tau]_{ab} $.
Moreover, it controls the number of samples that our method requires in general, as stated in the following.

\begin{thm}
	Let $ \rho, \sigma $ be $ n $-qubit states and $\{\hat\tau_i\}_{i=1}^N $ be $ N $ realisations of $[\hat \tau]_{ab}$, with $ a\sim P_\rho , b\sim P_\sigma$. Then $| \frac1N \sum_i \hat\tau_i - \tr(\rho\sigma)| < \epsilon $ with probability $ 1-\delta $ whenever $ N > \frac{\nu^n}{2\epsilon^2}\log\left(\frac{2}{\delta}\right) $.
\end{thm}

The proof follows directly from Hoeffding's inequality.
Hence, we see the sample complexity scales exponentially on the number of qubits according to $\nu$, which encodes the sample complexity overhead inherent to the quasiprobabilictic representation.

\section{Numerical experiments}

We performed numerical experiments to validate our estimator by taking $ n $ tensor copies of the 6-outcome POVM 
\begin{equation}\label{povm}
\{M_a = \frac13\ket{a}\bra{a}\}_a,
\end{equation}
where $ \ket{a}$ is an eigenvector of either $ X, Y,$ or $ Z $, and $\tau_1=\tau_2=T^+$, where $ T^+ $ denotes the pseudoinverse of $ T $.
Due to the freedom in the choice per equation Eq. (\ref{tau}), we considered different POVMs for our experiments.
We performed a brute-force search on the 4-, 6-, and 8-outcome qubit POVMs by incrementally varying the angles between Bloch-vectors representing the POVM elements.
This revealed the Pauli-6 POVM as the optimal choice.
We also used a Markov chain Monte Carlo\cite{Metropolis1953mh, Sherlock2010rwmh} dynamics to simulate a random walk on the set of solutions to Eq. (\ref{tau}) for minimising $ \nu $.
However, both searches provided only worse negativities.
This is further detailed in Ref. \cite{roeland}.

\begin{figure}[t]
	\includegraphics[width=\linewidth]{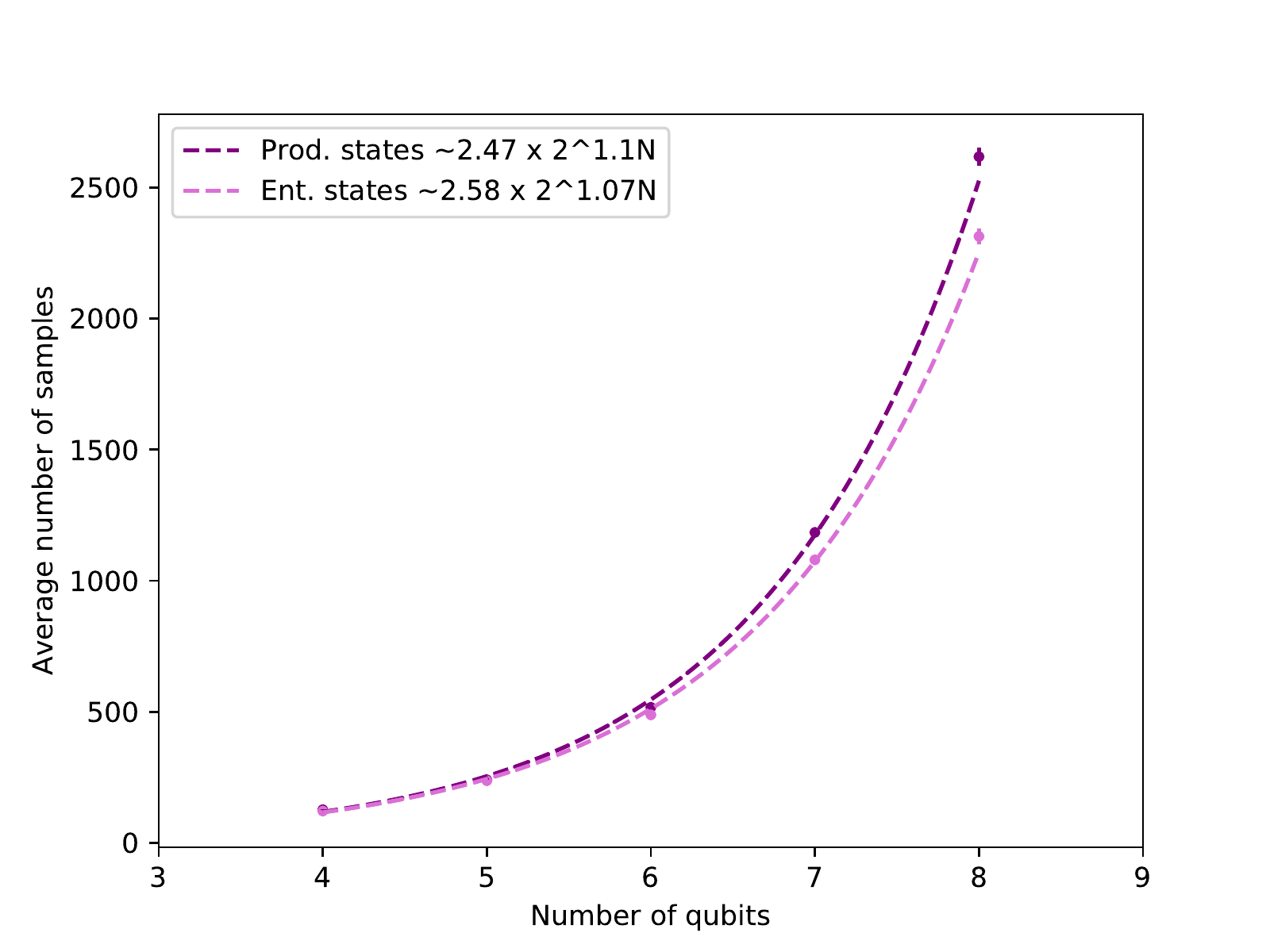}
	\caption{Average number of samples required for the absolute mean error of the overlap estimation to be $ <0.05 $, considering 180 different pairs of either product or entangled pure random states, according to the number of qubits in the states.}
	\label{fig:scaling}
\end{figure}

We numerically investigate the number of samples $ N $ required to make the estimation's absolute error, $ |\tr(\rho\sigma) - \frac1N\sum_i\hat\tau_i| $, lower than a fixed threshold.
We estimate the overlap of pure product and entangled pairs of states, constructed as matrix-product states from random local tensors.
In agreement with Ref. \cite{Elben2020}, we find that the latter is a slightly simpler problem.
In Fig. \ref{fig:scaling}, we present the obtained scaling for the fixed error bound of 0.05.
To minimise numerical fluctuations, we check the number of samples required to make the average error of 5 random pairs lower than this bound, for 100 batches of 5 states.
For more details, we refer to our computational appendix \cite{comp_app}.

\section{Connection with previous approaches}

We now investigate other overlap estimators, which we simulate in order to compare them with the quasiprobabilistic method. 
These alternative methods are separated in two groups.
We start from the ones that require coherently-controlled  operations between the two systems, described as quantum circuits.
Next, we consider two direct estimators based on random measurements and the theory of unitary $ t $-designs. 

\textbf{Entanglement-based approaches.---}
There are various quantum circuits designed specifically for calculating the overlap between quantum states.
However, considering realistic implementations, most NISQ devices have a series of restrictions in common.
For instance, the number of implementable quantum gates is limited, as is the connectivity between different qubits.
In addition, although the single-qubit gate noise is usually negligible, the noise present in multiqubit gate implementations cannot be ignored.
To address these points in our simulations of noisy circuits, we used the locally purified density operator (LPDO) paradigm, designed for simulating non-unitary evolutions in tensor networks language (see Appendix \ref{app:lpdo}).

Based on that, we benchmark our method considering quantum computers with the following features: (i) the set of implementable gates comprises noiseless single-qubit gates and noisy CNOTs; (ii) the circuits are one-dimensional; and (iii) the interactions between the qubits are limited to nearest-neighbours, meaning that long-distance CNOTs have to be compiled via first-neighbour CNOTs. 
The optimal compilation then adds $4(d-1)$ extra CNOTs to the original circuit, where $ d $ is the distance between the qubits in the long-distance CNOT.
We assume the noisy CNOTs are well modelled by an ideal CNOT followed by a depolarising channel.
This yields a circuit structure with increasing infidelity.
We emphasise the one-dimensional structure is considered only for simplicity and illustrative purposes.
Our conclusions hold also for 2D structures, but the corresponding classical simulation is more expensive.

Under these conditions, let us recall three circuits for state overlap.
The first is the standard SWAP test\cite{Buhrman2001}, which requires a single-qubit measurement on the ancillary qubit. 
Similarly, the improved SWAP test circuit\cite{Cincio2018} optimises the compilation of the standard circuit in view of constraint (i), demanding a lower number of layers and CNOTs but maintaining the single-qubit measurement on the ancilla.
Finally, on a different line, the Bell-basis circuit\cite{Garcia2013} requires a global measurement on the Bell basis in exchange of reducing the number of gates and layers.
With the limited connectivity assumption (iii), these circuits already involve a high number of CNOT gates for a small number of qubits (see Table \ref{tablee}).

\begin{table}[h]
	\begin{tabular}{c|c|c|c|c|c|}
		\cline{2-6}
		\textbf{}                           & 1Q & 2Q & 3Q  & ... & 8Q   \\ \hline
		\multicolumn{1}{|c|}{Improved SWAP test} & 12 & 60 & 144 & ... & 1104 \\ \hline
		\multicolumn{1}{|c|}{Bell-basis circuit}    & 1  & 10 & 27  & ... & 232  \\ \hline
	\end{tabular}
	\caption{Number of nearest-neighbours CNOTs required from the improved SWAP test and Bell-basis circuits, according to the number of qubits in the states.}
	\label{tablee}
\end{table}

\begin{figure}[t]
	\centering
	\includegraphics[width=0.9\linewidth]{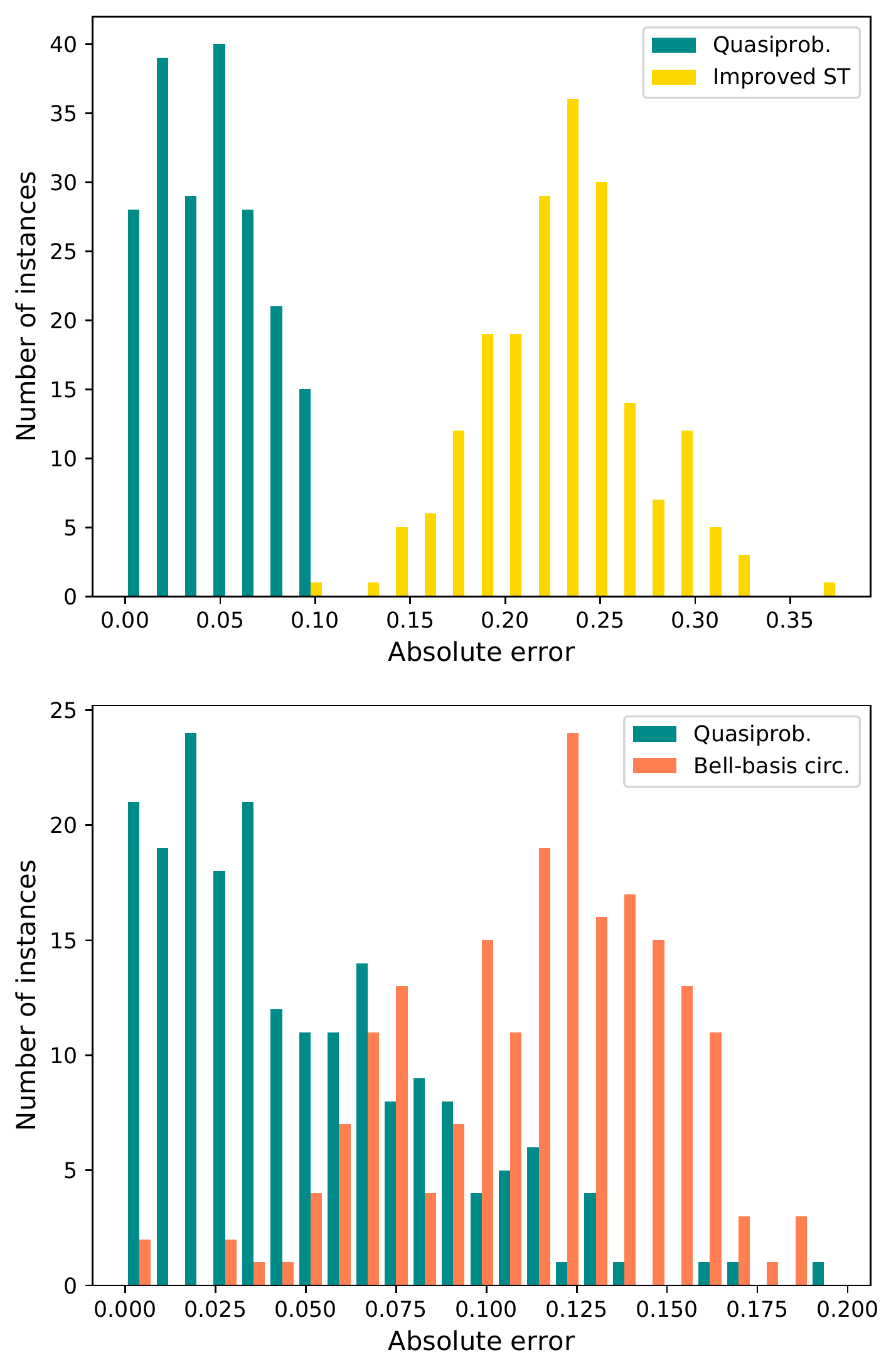}
	\caption{Histograms of the absolute error in the overlap estimation of 180 pairs of states by the quasi-probabilistic and circuit-based methods. In both circuits we consider a CNOT-noise of 0.005 (single-qubit gates are implemented noiselessly) and measurement noise of 0.01. (a) The improved SWAP-test. The comparison is done based on 500 samples from two-qubit random states (with an averaged overlap of 0.86), for which the near-neighbours circuit has 54 layers. (b) The Bell-basis test. We use 1,500 samples from 3-qubit random states (with an averaged overlap of 0.78), for which the near-neighbours circuit has 28 layers.}
	\label{fig:circuits}
\end{figure}

In these approaches, on top of the statistical error of estimating the expectation value of an observable, we have the systemic noise arising from the implementation of two-qubit gates.
In contrast, using the quasi-probabilistic direct approach we obtain a qualitative advantage since the estimation (\ref{estimator}) can be as accurate as needed, as long as the number of samples is large enough.
As shown in Fig. \ref{fig:circuits}, even for a modest number of samples we already see a clear advantage to our method in the cases of 2- and 3-qubit states.

\begin{figure}[t]
	\includegraphics[width=\linewidth]{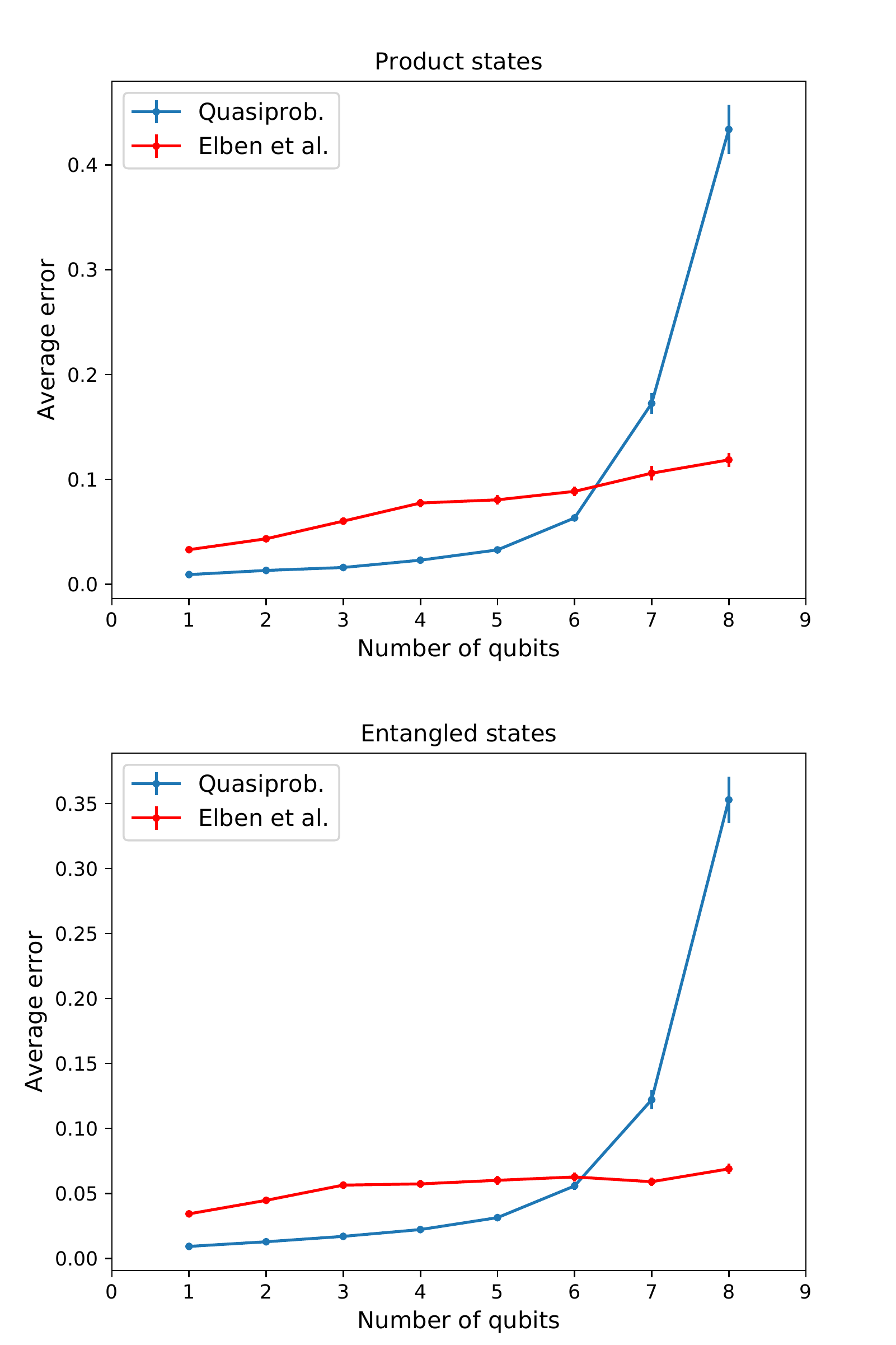}
	\caption{Comparison between the quasi-probabilistic and Elben et al. estimator\cite{Elben2020}, for a fixed budget of 10,000 samples, for random pure product and pure entangled states. In the latter, we used 100 samples from each of 100 random measurements. The mean absolute error was computed over 200 instances and the average overlap was $\approx0.6$ in both cases.}
	\label{fig:inn}
\end{figure}

\textbf{Random unitary-based methods.---}
We now compare our estimator with the two state-of-the-art direct techniques for calculating state overlaps.
The first technique\cite{Elben2020} applies the same random measurements on both states, coordinated by means of classical communication.
Nevertheless, it is also entanglement-free and allows for independent interaction with each system.
Next, it uses the theory of unitary $ t $-designs for  approximating the overlap.
The direct performance comparison with our method (Fig. \ref{fig:inn}) shows that our approach presents a lower mean absolute error for $ n $-qubit states, with $ n<7 $.
However, our method is outperformed for states involving more qubits.
This is confirmed by the sample complexity required to obtain an error bound of $ 0.05 $, $ \sim 2^{1.1n}, $ shown in Fig. \ref{fig:scaling}, while the reported corresponding scaling of the random measurements method is $ \sim 2^{0.8n} $.

Finally, we discuss the connection of our results with shadow estimation\cite{Huang_2020, kliesch2021}, a recently proposed method to efficiently estimate linear functions of quantum states.
The so-called classical shadow of a state $ \rho $, given by the inverse of the map $ \mathcal{M}(\rho) = \mathbb{E}_{U, b}(U\ket b\bra b U^\dagger) $, is obtained by taking the average over some ensemble of unitaries $ U $ to $ \rho $ and subsequently measuring on the computational basis $ \{\ket b\bra b\}_b $.
Whenever the unitaries are chosen from the local Clifford group, the corresponding shadow fidelity estimator matches Eq. (\ref{estimator}), for $ M_a^{(1)}$ as in Eq. (\ref{povm}) and for a particular choice of $ \tau_1, \tau_2 $ (see Appendix \ref{app:shadow}).
Therefore, these specific instances of each estimator share the same performance\cite{Huang_2020}.

\section{Final discussion}

While efficient entanglement-based quantum algorithms for approximating the overlap of quantum states exist, the currently available devices imply unavoidable systematic errors.
Therefore, direct and entanglement-free methods may provide an advantage in the estimation of state overlaps in the near term.
These typically present only statistical errors and treat the overlap estimation as a sample-complexity problem.
Among the entanglement-free direct estimators, we show that our method presents the lowest mean absolute error for n-qubits states, with n < 7.
We also identify that the quasiprobabilistic construction shares some common elements with the shadow fidelity estimation approach.
Moreover, for particular choices on each of the methods, the constructed estimators are the same.

Our work opens up avenues for future explorations.
In here, we have used a quasiprobabilistic representation
specifically for estimating the fidelity between quantum
states.
However, these representations can be used to describe not only quantum states, but also quantum measurements, unitary evolution, and open system quantum dynamics\cite{Pashayan2015, Carrasquilla_21, Luo_21, Reh_21}.
Can they be efficiently used for classical simulations of quantum circuits?
For what circuits the associated sample-complexity overhead becomes too expensive?
It would be interesting to investigate the interplay between these figures in the context of classical-quantum simulations.

\textit{Acknowledgements} ---
The authors thank Ingo Roth for helpful discussions.
LG and LA were financially supported by the Serrapilheira Institute (grant number Serra-1709- 17173) and the Brazilian agencies CNPq (PQ grant No. 311416/2015-2 and INCT-IQ), FAPERJ (PDR10 E- 26/202.802/2016 and JCN E-26/202.701/2018) and S\~ao Paulo Research Foundation (FAPESP) under grants 2016/01343-7 and 2018/04208-9.
LA also acknowledges support from the Technology Innovation Institute of Abu Dhabi, United Arab Emirates.
JC acknowledges support from the Natural Sciences and Engineering Research Council of Canada (NSERC), the Shared Hierarchical Academic Research Computing Network (SHARCNET), Compute Canada, Google Quantum Research Award, and the CIFAR AI chair program.
Resources used in preparing this research were provided, in part, by the Province of Ontario, the Government of Canada through CIFAR, and companies sponsoring the Vector Institute www.vectorinstitute.ai/\#partners.

\begin{appendices}

\section{Proof of Lemma \ref{lemma}}\label{app:lemma}

In the main text, we showed that given an informationally-complete POVM $ \{M_a\} $ and its associated matrix $ [T_{ij}]_{ij} = \tr(M_iM_j) $, if the representation $ \rho = \sum_{a,a'} P_\rho(a)\tau_{aa'}M_{a'} $ holds for any quantum state, then $ \tau $ satisfies $ T = T\tau T $.
Here we will show the converse, completing the proof of Lemma \ref{lemma}.

From completeness, we can write any state as $ \rho = \sum_i \alpha_i M_i $.
Hence, for any $ M_j $, we have
\begin{equation*}\label{key}
\tr(\rho M_j) = \sum_i \alpha_i \tr(M_iM_j) = \sum_i \alpha_i T_{ij}.
\end{equation*}
Assuming $ T = T \tau T $, we obtain
\begin{align*}\label{key}
\tr(\rho M_j) =& \sum_i \alpha_i \left(\sum_{a, a'}T_{ia}\tau_{aa'}T_{a'j}\right)\\
=& \sum_{a, a'} \left(\sum_i \alpha_i T_{ia}\right)\tau_{aa'}T_{a'j}\\
=& \sum_{a,a'}\tr\left(\left(\sum_i\alpha_iM_i\right) M_a\right)\tau_{aa'}\tr(M_{a'}M_j)\\
=& \sum_{a,a'}P_\rho(a)\tau_{aa'}\tr(M_{a'}M_j).
\end{align*}
Since this holds for all $ j $, we conclude
\begin{equation}\label{key}
\rho = \sum_{a,a'}P_\rho(a)\alpha_{aa'}M_{a'}.
\end{equation}

\section{The LPDO paradigm}\label{app:lpdo}

Numerical simulations with the full density matrix of size $2^n\times 2^n$ quickly become prohibitive due to the large memory requirements.
Hence, in order to compare these approaches efficiently, we describe our quantum states as locally purified density operators (LPDOs), within the tensor networks paradigm.
This framework is particularly useful since we perform a simulation of \textit{noisy} quantum circuits, in which the implementation of each CNOT is followed by the performance of a depolarising channel on the respective qubits.
Hence, the updated circuit state is mixed.

The canonical choice for representing operators with tensor networks are matrix product operators (MPO) \cite{Verstraete2004mpo}.
A drawback of this approach is that applying completely positive maps to the state can still lead to the MPO becoming non-positive due to truncation errors.
The locally purified density operator method solves this issue by representing the state as $\varrho = XX^\dag$, where the purification operator $X$ is given by a tensor network
\begin{align}
[X]^{p_1, \ldots, p_n}_{k_1, \ldots, k_n} = \sum_{b_1,\ldots,b_{n-1}} A^{[1] p_1, k_1}_{b_1} A^{[2] p_2, k_2}_{b_1,b_2}\ldots  A^{[l] p_n k_n}_{b_{n-1}},
\end{align}
where $1\leq p_l \leq P$, $1\leq k_l \leq K$ and $1\leq b_l \leq D$.
Here, $P$ is called the physical dimension, $K$ is the Kraus dimension and $D$ is the bond dimension. Hence we can guarantee the positivity of our tensor network at the cost of introducing an additional dimension $k_l$ into our local tensors $A^{[l]}$.

In general, applying an entangling gate increases the cardinality of $ b_l $, while applying a depolarising channel increases the cardinality of $ k_l $.
In order to keep the simulation efficient, we must ensure the bounds $ D $ and $ K $, called bond dimension and Kraus dimension, respectively, are respected.
To do so, whenever these bounds are exceeded we truncate the corresponding description via a singular value decomposition on the respective tensors; for truncations on the Kraus dimension, the LPDO formalism guarantees that the final description refers to a positive semi-definite operators and our description of the circuit remains valid.

We can control the accuracy of the simulation by increasing $D$ and $K$ and keeping track of a runtime lower bound estimate of the state fidelity.
Let $\varrho = X^\dag X$ and $\sigma=Y^\dag Y$.
Then the fidelity is given by
\begin{align}
F(\varrho, \sigma) = \tr{\sqrt{\sqrt{\sigma}\varrho\sqrt{\sigma}}}.
\end{align}
From Lemma 1 in \cite{Werner2016positivetensor} we know that, 
\begin{align}
F(\varrho, \sigma) \geq \frac{1}{2}\left(2 - \norm{X - Y}_2^2\right).
\end{align}
Let $X$ be a locally purified description of a quantum state with local tensors $\{A^{[n]}\}$ that is in mixed canonical form with respect to a local tensor $A^{[l_{cp}]}$. If a single tensor $A^{[n_{cp}]}$ is compressed by a discarding singular values in either the Kraus or bond dimensions the error, then by lemma 6 in \cite{Werner2016positivetensor} we know that
\begin{align}
\delta := \left(\sum_{i, \text{discarded}} s_i^2 \right)^{\frac{1}{2}},
\end{align}
and subsequently
\begin{align}
\norm{X - X'}_2^2 = 2(1 - \sqrt{1 - \delta^2})
\end{align}
where $X'$ is the compressed tensor.
By the triangle inequality, the two norm errors introduced by the discarded weights can at most sum up. Hence the true operator norm is lower bounded by the sum of all discarded weight errors
\begin{align}
\norm{\varrho_{\text{exact}} - \varrho_{\text{truncated}}}_2 \leq \sum_{d}  \sqrt{2(1 - \sqrt{1 - \delta_d^2})}
\end{align}
With $d$ the number of trunctations and $\delta_k$ the discarded weights.
This brings the final runtime fidelity estimate to
\begin{align}
F(\varrho, \sigma) & \geq \frac{1}{2}\left(2 - \norm{X - Y}_2^2\right) \\
&\geq \frac{1}{2}\left(2 - \left(\sum_{d}  \sqrt{2(1 - \sqrt{1 - \delta_d^2})}\right)^2\right)\label{eq:bound}
\end{align}
In all our noisy experiments, we apply depolarizing channels only to the two qubit gates, since single qubit gate noisy tends to be small in experimental settings. The channel is given by
\begin{align}
\mathcal{\varrho} = \sum_{m=1}^M K_m \varrho K_m^\dag,
\end{align}
where $\{K_m\}$ is a set of Kraus operators with
\begin{align}
K_1 &= \sqrt{\frac{4 - 3 \lambda}{4}}  I,\quad
K_2 = \sqrt{\frac{\lambda}{ 4}} \sigma^x\\
K_3 &= \sqrt{\frac{\lambda}{ 4}} \sigma^y, \quad
K_4 =  \sqrt{\frac{\lambda}{ 4}} \sigma^z,
\end{align}
where $\lambda\in[0,1]$ is the depolarising factor. 

The tensor networks formalism is also convenient for describing our quasi-probabilistic method, since the involved POVM and its corresponding $ \tau $ matrix are factorable and act locally, so the tensor contractions are done most efficiently.
For more more details on our implementations, please see our computational appendix \cite{comp_app}.

\section{Explicit connection to shadow estimation}\label{app:shadow}

Our method requires an IC-POVM and at least one choice of a generalised inverse for T, i.e. a matrix $ \tau $ satisfying $ T=T\tau T $, where $ T $ is the associated matrix of the POVM.
We used the Pauli-6 POVM (Eq. (\ref{povm})) and $ \tau  = T^+$, the pseudoinverse of $ T $; however, there are infinitely many matrices that satisfy Eq. (\ref{tau}).

On the other hand, the classical shadows technique\cite{Huang_2020} uses unitaries $ \{U_i\} $, randomly chosen from a distribution $ \mu $, and measurements on the computational basis $ \ket{a} =\ket0, \ket1$ to reconstruct any $ n $-qubit state $ \rho $ as
\begin{equation}\label{key}
\rho = \sum_{i, a} \mu(i)\bra a U_i \rho U_i^\dagger \ket a \left[(2^n+1)U_i\ket a \bra a U_i^\dagger - \mathbb{I}\right].
\end{equation}
For $ n=1 $, setting $ \mu $ as the uniform distribution over the Pauli unitaries $ U_i = X, Y, Z $ and combining them with the computational basis as $ M_{i, a} = U_i\ket a \bra a U_i^\dagger$, we obtain
\begin{align}\label{key}
\rho &= \sum_{i, a} \frac13 \tr(M_{i, a}\rho)\left[3M_{i, a} - \mathbb{I}\right]\\
&= \sum_{i, a} \frac13 \tr(M_{i, a}\rho)\left[2M_{i, a} - M_{i, a'}\right],
\end{align}
where $ a' = a+1 \mod 2 $ and we use that fact that $ \mathbb{I} = M_{i, 0}+M_{i, 1} $ for $ i=1,2,3 $.
Hence, we can rewrite the square bracket as
\begin{equation}\label{key}
2M_{i, a} - M_{i, a'} = \sum_{b}\tilde\tau_{ab}M_{i, b},
\end{equation}with
\begin{equation}\label{taushadow}
\tilde\tau = \begin{bmatrix}
	2& -1& 0& 0& 0& 0 \\
	-1& 2& 0& 0& 0& 0 \\
	0& 0& 2& -1& 0& 0 \\
	0& 0& -1& 2& 0& 0 \\
	0& 0& 0& 0& 2& -1 \\
	0& 0& 0& 0& -1& 2 \\
\end{bmatrix}.
\end{equation}
\end{appendices}
We finish recognising the final expression as a quasiprobabilistic representation by rescaling the operators $ M_{i, a} $ using the $ 1/3 $ factor to obtain the elements of the Pauli-6 POVM (which leads to a rescaling $ \tilde\tau \mapsto 3\tilde\tau$).

These observations were our motivation to investigate the plurality of $ \tau $'s and its optimisations, that lead to Lemma \ref{lemma}.
However, as reported in the main text, we concluded that the pseudoinverse $ T^+ $ is the best option.
Indeed, it is actually equivalent to $ \tilde\tau $, as one can check that $ \tilde\tau T = T^+T $, and therefore the relevant tensor for our overlap estimation method, $ \tau T\tau $, is the same in both cases.

\end{document}